\documentclass[aps,prd,twocolumn,superscriptaddress,nofootinbib]{revtex4-1}

\usepackage{latexsym}
\usepackage{amsmath}
\usepackage{amssymb}
\usepackage{amsfonts}
\usepackage{bm}
\usepackage{physics}
\usepackage[table,x11names]{xcolor}
\usepackage{color}
\definecolor{purple}{rgb}{0.5,0,0.5}
\definecolor{blue}{rgb}{0.0,0,0.9}
\definecolor{prdblue}{rgb}{0.133,0.118,0.498}
\usepackage[colorlinks=true, pdfstartview=FitV, linkcolor=prdblue, citecolor= prdblue, urlcolor=prdblue]{hyperref}

\usepackage{supertabular} 
\usepackage{placeins}
\usepackage{epsfig}
\usepackage{graphicx}

\def\tstrut{\vrule height3.25ex depth0pt width0pt} 

\begin{document}


\title{Exploring $T_{\Upsilon\Upsilon}$ tetraquark candidates in a coupled-channels formalism}


\author{P.G. Ortega}
\email[]{pgortega@usal.es}
\affiliation{Departamento de F\'isica Fundamental, Universidad de Salamanca, E-37008 Salamanca, Spain}
\affiliation{Instituto Universitario de F\'isica 
Fundamental y Matem\'aticas (IUFFyM), Universidad de Salamanca, E-37008 Salamanca, Spain}

\author{D.R. Entem}
\email[]{entem@usal.es}
\affiliation{Departamento de F\'isica Fundamental, Universidad de Salamanca, E-37008 Salamanca, Spain}
\affiliation{Instituto Universitario de F\'isica
Fundamental y Matem\'aticas (IUFFyM), Universidad de Salamanca, E-37008 Salamanca, Spain}

\author{F. Fern\'andez}
\email[]{fdz@usal.es}
\affiliation{Instituto Universitario de F\'isica 
Fundamental y Matem\'aticas (IUFFyM), Universidad de Salamanca, E-37008 Salamanca, Spain}

\author{J. Segovia}
\email[]{jsegovia@upo.es}
\affiliation{Departamento de Sistemas F\'isicos, Qu\'imicos y Naturales, Universidad Pablo de Olavide, E-41013 Sevilla, Spain}

\date{\today}

\begin{abstract}
We investigate the spectrum of $T_{\Upsilon\Upsilon}$ tetraquark candidates within a coupled-channels framework. The analysis includes all $L\leq2$ combinations of $\Upsilon(1S)$, $\Upsilon(2S)$, $\eta_b(1S)$, and $\eta_b(2S)$ in the $J^P = 0^\pm, 1^\pm, 2^\pm$ sectors. The meson-meson interaction is derived from an underlying constituent quark model through the resonating group method, and the properties of the states are obtained from poles of the scattering matrix.
We find a rich spectrum of resonant, and virtual, states distributed between the $\eta_b(1S)\eta_b(1S)$ and $\Upsilon(2S)\Upsilon(2S)$ thresholds. The pattern of poles exhibits approximate heavy-quark spin symmetry multiplets. Several states are dominated by a single channel and can be associated with threshold-driven structures, while higher-mass resonances show sizable mixing among channels involving radially excited bottomonia. The predicted widths range from tens to several hundred MeV.
Branching ratios indicate that many states couple predominantly to final states with at least one excited bottomonium, whereas only a subset of the spectrum is expected to be visible in the $\eta_b(1S)\eta_b(1S)$, $\eta_b(1S)\Upsilon(1S)$ and $\Upsilon(1S)\Upsilon(1S)$ channels. These results provide quantitative guidance for experimental searches of fully heavy tetraquarks and offer a test of coupled-channel dynamics and heavy-quark spin symmetry in the $bb\bar b\bar b$ sector.
\end{abstract}


\maketitle


\section{INTRODUCTION}
\label{sec:introduction}

The study of hadronic states beyond the conventional quark-antiquark and three-quark configurations remains a central subject in hadron spectroscopy~\cite{ParticleDataGroup:2024cfk}. In recent years, a number of exotic candidates that cannot be easily accommodated within the naive quark model~\cite{GellMann:1964nj, Zweig:1964CERN} have been reported experimentally, stimulating extensive theoretical and phenomenological efforts to understand their nature~\cite{Dong:2020hxe, Chen:2016qju, Chen:2016spr, Guo:2017jvc, Liu:2019zoy, Yang:2020atz, Dong:2021bvy, Chen:2021erj, Cao:2023rhu, Mai:2022eur, Meng:2022ozq, Chen:2022asf, Guo:2022kdi, Ortega:2020tng, Huang:2023jec, Lebed:2023vnd, Zou:2021sha, Du:2021fmf, Liu:2024uxn, Johnson:2024omq, Entem:2025bqt, Hanhart:2025bun, Wang:2025dur, Wang:2025sic, Francis:2024fwf, Chen:2024eaq, Husken:2024rdk, Liu:2024uxn, Johnson:2024omq}. A particularly intriguing class of such exotic states are the fully heavy tetraquarks, composed solely of heavy quarks and antiquarks, $Q\bar Q Q\bar Q$ with $Q=c,b$. 

On the experimental front, the LHCb collaboration reported structures in the di-$J/\psi$ invariant mass spectrum in proton-proton collisions, including a relatively narrow peak around $6.9$ GeV and broader enhancements at lower and higher masses, consistent with fully charmed tetraquark candidates~\cite{LHCb:2020bwg}. Subsequent measurements by the CMS and ATLAS collaborations have confirmed these observations, indicating a rich structure in the fully-charm sector~\cite{CMS:2023owd, ATLAS:2022hhx}. These discoveries have revitalized interest in fully heavy multiquark systems and provided concrete experimental targets for theoretical models.

In contrast, experimental evidence for fully bottom tetraquarks $(bb\bar b\bar b)$ is still lacking. Searches for resonances in the $\Upsilon(1S)\Upsilon(1S)$ spectrum have so far yielded no significant signals, although the accessible phase space and production mechanisms at the LHC make such searches challenging~\cite{LHCdesign}. Nonetheless, experimental efforts continue, and future high-luminosity data sets are expected to substantially improve sensitivity~\cite{Apollinari:2015bam}.

The theoretical landscape for fully heavy tetraquarks has been extensively explored using a variety of complementary approaches. Early quark-model studies of fully heavy systems predicted compact tetraquark configurations~\cite{Berezhnoy:2011xn, Karliner:2016zzc, Wu:2016vtq, Anwar:2017toa, Debastiani:2017msn, Eichten:2017ffp, Deng:2020iqw, Yang:2021hrb, Asadi:2021ids, Wu:2025nbu}. In the fully-bottom case, masses are typically found in the $18-19$~GeV region. QCD sum rule analyses have provided independent estimates of the masses and quantum numbers of fully bottom tetraquark candidates, with results broadly consistent with quark model expectations but also highlighting significant theoretical uncertainties~\cite{Chen:2016jxd, Wang:2017jtz, Wang:2018poa, Wang:2019rdo, Zhang:2020xtb}. Within effective field theory frameworks, the role of heavy-quark spin symmetry (HQSS) and threshold effects has been studied to elucidate structure and multiplet patterns in the fully heavy sector~\cite{Mehen:2015gra, Guo:2018kno}. Recent lattice QCD studies have reported mixed conclusions, with some indications of near-threshold attraction~\cite{Hughes:2017xie}, while others do not find deeply bound compact states~\cite{Junnarkar:2019equ}. More recent coupled-cluster, diffusion Monte Carlo, and other many-body methods have further enriched the theoretical picture, offering complementary insights into the dynamics of four-heavy-quark systems~\cite{Barnea:2006sd, Lloyd:2013bza, Bai:2016int, Richard:2017vry, Bedolla:2019zwg, Liu:2019zuc, Gordillo:2020sgc}. Beyond static mass predictions, production mechanisms of fully heavy tetraquarks in hadronic collisions have been addressed in perturbative QCD frameworks, estimating cross sections and kinematic distributions relevant for LHC experiments~\cite{Esposito:2018cwh, liu:2020eha, Belov:2024qyi}.

A widely-used constituent quark model based meson-meson coupled-channels calculation have been recently applied to the fully charm tetraquark spectrum~\cite{Ortega:2023pmr}. In particular, meson-meson interactions derived through the resonating group method (RGM) in combination with a scattering-matrix pole analysis, provided a unified description of several $T_{\psi\psi}$ candidates structures reported experimentally. That study established a theoretical framework in which resonant and virtual states arise naturally from the interplay between nearby thresholds and short-range quark-exchange dynamics.

In the present work, we extend this framework to the fully bottom sector by investigating the spectrum of $T_{\Upsilon\Upsilon}$ candidates generated by the interaction of bottomonium mesons, including $\eta_b(1S)$, $\eta_b(2S)$, $\Upsilon(1S)$ and $\Upsilon(2S)$. Using the same coupled-channels formalism within the constituent quark model and identifying physical states as poles of the scattering matrix, we systematically explore the energy region spanning from the lowest threshold up to the $\Upsilon(2S)\Upsilon(2S)$ one. This approach enables a direct and consistent comparison with the results obtained in the fully charmed sector, while yielding concrete predictions for the masses, widths, and dominant decay channels of fully bottom tetraquark candidates that may be accessible in future high-luminosity experiments.

Although a large number of studies have addressed the fully-bottom sector, the theoretical situation is far from settled. While many approaches predict structures in the $18-19$~GeV region, the dynamical origin of these states remains model dependent. In particular, some lattice QCD calculations do not find clear evidence for deeply bound compact tetraquarks, and several quark-model analyses incorporate explicit color-octet components, which enhance the role of confinement dynamics at short distances.

In this context, it is important to clarify to what extent the observed structures can be generated solely from the interaction between physical color-singlet bottomonia, once quark-exchange effects are consistently included. The coupled-channels framework employed here provides precisely such a test: it derives the meson-meson interaction from an underlying constituent quark model via the resonating group method, but restricts the basis to asymptotic color-singlet meson configurations. Therefore, any resonance that appears is generated dynamically from quark rearrangement effects rather than from explicit compact diquark-antidiquark or color-octet configurations.

The present calculation is thus complementary to compact tetraquark approaches. It addresses the question of whether the structures predicted in the literature can be interpreted as threshold-driven states emerging from coupled-channel dynamics, and it provides quantitative predictions for pole positions, widths, and branching ratios within a unified framework already tested in the fully-charmed sector.

The organization of the manuscript is as follows: After this introduction, Section~\ref{sec:theory} provides a brief overview of the theoretical framework. Section~\ref{sec:results} primarily focuses on the analysis and discussion of our theoretical findings. Lastly, in Sec.~\ref{sec:summary}, we present a summary of our work and draw conclusions based on the obtained results.


\section{THEORETICAL FORMALISM}
\label{sec:theory}

In this work, we investigate the $T_{\Upsilon\Upsilon}$ tetraquark candidates as meson-meson molecular configurations of $\Upsilon(1S)$, $\Upsilon(2S)$, $\eta_b(1S)$ and $\eta_b(2S)$ in the $bb\bar b\bar b$ system. The theoretical framework is identical to the one previously employed for the fully charmed sector~\cite{Ortega:2023pmr}, and here we summarize only those features that are relevant for the present bottomonium-bottomonium study.

\subsection{Constituent quark model}

The underlying quark--(anti-)quark interaction is described within a constituent quark model (CQM)~\cite{Vijande:2004he} that has been extensively applied to hadron spectra, hadron-hadron interactions, and multiquark systems across a wide range of energies~\cite{Fernandez:1993hx, Valcarce:1994nr, Valcarce:1995dm, Vijande:2006jf, Segovia:2008zza, Segovia:2008zz, Segovia:2009zz, Ortega:2009hj, Segovia:2011zza, Segovia:2015dia, Yang:2015bmv, Ortega:2016hde, Ortega:2016mms, Ortega:2016pgg, Yang:2017qan, Yang:2017rpg, Ortega:2018cnm, Ortega:2020uvc, Yang:2020atz}. The model incorporates three main contributions: (i) an effective one-gluon exchange interaction, accounting for short-range QCD dynamics; (ii) a color confinement, representing the non-perturbative long-range interaction; and (iii) Goldstone-boson exchanges, relevant only for light-light quark pairs. Since the present system involves only heavy quarks, chiral symmetry is explicitly broken and Goldstone-boson exchanges are absent. Therefore, the interaction is driven exclusively by one-gluon exchange and color confining terms.

Regarding color confinement, while it has been proven that multi-gluon exchanges generate an attractive potential that rises linearly with the distance between infinitely heavy quarks~\cite{Bali:2000gf}, sea quarks contribute to screening the rising potential at low momenta and eventually lead to the breaking of the quark-antiquark binding string~\cite{Bali:2005fu}. To account for this behavior, the confinement potential is written as
\begin{equation}
V_{\text{CON}}(\vec r\,)=\Big[-a_c\big(1-e^{-\mu_c r}\big)+\Delta\Big]
(\vec\lambda^c_i\cdot\vec\lambda^c_j) \,,
\end{equation}
where $\vec{\lambda}^c$ are the color Gell-Mann matrices along with $a_c$, $\mu_c$, and $\Delta$ that are model parameters. At short distances the potential behaves linearly, while it saturates at large distances.

The perturbative contribution arises from the effective one-gluon exchange (OGE) interaction derived from the Lagrangian
\begin{equation}
\mathcal L_{qqg}=i\sqrt{4\pi\alpha_s} \, \bar\psi \gamma^\mu G_\mu^c \lambda^c \psi \,,
\end{equation}
where $\alpha_s$ is an effective scale-dependent strong coupling constant,
\begin{equation}
\alpha_s(\mu)=
\frac{\alpha_0}
{\ln\!\left[\left(\mu^2+\mu_0^2\right)/\Lambda_0^2\right]} \,,
\end{equation}
with $\mu$ the reduced mass of the interacting quark pair, along with $\alpha_{0}$, $\mu_{0}$ and $\Lambda_{0}$ that are model parameters. Final expression of the potential is given by  
\begin{align}
V_{\text{OGE}}(\vec{r}\,) &= \frac{1}{4} \alpha_{s} (\lambda_{q}^{c}\cdot \lambda_{\bar q}^{c}) \Bigg[ \frac{1}{r} \nonumber \\
& 
- \frac{1}{6m_{q}m_{\bar q}} (\vec{\sigma}_{q}\cdot\vec{\sigma}_{\bar q}) \frac{e^{-r/r_{0}(\mu)}}{r r_{0}^{2}(\mu)} \Bigg] \,,
\end{align}
where $\vec{\sigma}$ are the Pauli matrices. The smearing parameter $r_0(\mu)=\hat{r}_0/\mu$ depends on the reduced mass, $\mu$, of the interacting quark pair. In addition to the exchange diagrams mediated by a gluon, one can also include the quark--anti-quark interaction via gluon annihilation diagrams in the $S$-channel. In coordinate space, this interaction is given by~\cite{Faessler:1982qt, Entem:2006dt}:
\begin{align}\label{eq:OGEani}
V_{\text{Anh,OGE}}(\vec{r}\,) &= \frac{1}{(2\pi)^3} \, \delta^{(3)}(\vec{r}\,) \nonumber \\ 
&
\times \frac{\alpha_s}{8\pi^2m_{q}m_{\bar q}}
\left( \frac{4}{9} - \frac{1}{12} \, \vec \lambda_q^c \cdot \vec \lambda_{\bar q}^c \right)\nonumber \\ 
&
\times \left( \frac{3}{2} + \frac{1}{2} \, \vec \sigma_q \cdot \vec \sigma_{\bar q} \right) \left( \frac{1}{2} - \frac{1}{2} \, \vec \tau_q \cdot \vec \tau_{\bar q} \right) \,.
\end{align}

All model parameters are fixed from previous studies, so no new adjustments are introduced here. Their particular values can be consulted in Ref.~\cite{Segovia:2008zz}, which provides an updated parametrization of the original model~\cite{Vijande:2004he}. Besides, as in Ref.~\cite{Segovia:2016xqb}, the bottomonium ($b\bar b$) states are obtained by solving the two-body Schr\"odinger equation using the Gaussian Expansion Method~\cite{Hiyama:2003cu}, which provides accurate meson wave functions to be used as input in the meson-meson calculation.

It is worth emphasizing that, in the present formulation, the Hilbert space is restricted to products of physical color-singlet mesons. Hidden-color configurations do not appear as independent basis states. This choice allows us to isolate the role of coupled-channel dynamics among physical hadronic degrees of freedom.

\subsection{Meson-meson interaction: Resonating Group Method}

To describe the interaction between two bottomonium mesons, we employ the resonating group method (RGM)~\cite{Wheeler:1937zza}, which derives the meson-meson potential from the underlying quark dynamics.

The total wave function of a system composed of two bottomonia (A) and (B) is written as
\begin{equation}
\Psi = \mathcal A \big[ \phi_A \, \phi_B \, \chi_L \, \sigma_{ST} \, \xi_c \big] \,,
\end{equation}
where $\phi_{A(B)}$ are the internal wave functions of the mesons, $\chi_L$ describes their relative motion, $\sigma_{ST}$ is the spin-isospin wave function, and $\xi_c$ the color wave function. Since the system contains two identical $b$ quarks and two identical $\bar b$ antiquarks, the  anti-symmetrization operator,
\begin{equation}
\mathcal A=(1-P_b)(1-P_{\bar b}) \,,
\end{equation}
is needed. Note that $P_b$ and $P_{\bar b}$ are the exchange operators for quarks and antiquarks between clusters. Following Ref.~\cite{Ortega:2023pmr}, the antisymmetric operator reduces accordingly for identical mesons, while for nonidentical mesons both $AB$ and $BA$ configurations contribute.

The total interaction between the two color-singlet mesons can be separated into a direct kernel, with no quark exchange, and an exchange kernel, arising from quark rearrangement. The direct potential reads
\begin{align}
&
V_{D}(\vec{P}',\vec{P}_{i}) = \sum_{i\in A, j\in B} \int d\vec{p}_{A'} d\vec{p}_{B'} d\vec{p}_{A} d\vec{p}_{B} \times \nonumber \\
&
\times \phi_{A'}^{\ast}(\vec{p}_{A'}) \phi_{B'}^{\ast}(\vec{p}_{B'})
V_{ij}(\vec{P}',\vec{P}_{i}) \phi_{A}(\vec{p}_{A}) \phi_{B}(\vec{p}_{B}) \,,
\end{align}
where $V_{ij}$ is the CQM potential between the quark $i$ and the quark $j$ of the mesons $A$ and $B$, respectively. In the present system, direct contributions are small because confinement does not act between two color singlets, and gluon-annihilation diagrams, which are included in our calculations, are suppressed for heavy quarks. Thus, the dominant interaction arises from exchange terms. The exchange kernel can be written as
\begin{align}
K_E(\vec P',\vec P_i) &= H_E(\vec P',\vec P_i) - E_T\, N_E(\vec P',\vec P_i) \,,
\end{align}
which is a non-local and energy-dependent kernel, separated into a potential term $H_E$ plus a normalization term $N_E$. Here, $E_T$ denotes the total energy of the system and $\vec P_i$ is a continuous parameter. The exchange Hamiltonian and normalization can be written as
\begin{subequations}
\begin{align}\label{eq:exchangeV}
&
H_{E}(\vec{P}',\vec{P}_{i}) = \int d\vec{p}_{A'}
d\vec{p}_{B'} d\vec{p}_{A} d\vec{p}_{B} d\vec{P} \phi_{A'}^{\ast}(\vec{p}_{A'}) \times \nonumber \\
&
\times  \phi_{B'}^{\ast}(\vec{p}_{B'})
{\cal H}(\vec{P}',\vec{P}) P_{\bar b} \left[\phi_A(\vec{p}_{A}) \phi_B(\vec{p}_{B}) \delta^{(3)}(\vec{P}-\vec{P}_{i}) \right] \,,\\
&
N_{E}(\vec{P}',\vec{P}_{i}) = \int d\vec{p}_{A'}
d\vec{p}_{B'} d\vec{p}_{A} d\vec{p}_{B} d\vec{P} \phi_{A'}^{\ast}(\vec{p}_{A'}) \times \nonumber \\
&
\times  \phi_{B'}^{\ast}(\vec{p}_{B'})
P_{\bar b} \left[\phi_A(\vec{p}_{A}) \phi_B(\vec{p}_{B}) \delta^{(3)}(\vec{P}-\vec{P}_{i}) \right] \,,
\end{align}
\end{subequations}
where ${\cal H}$ denotes the Hamiltonian at quark level.

\subsection{Scattering equation and pole extraction}

The properties of the $T_{\Upsilon\Upsilon}$ candidates are obtained as poles of the meson-meson scattering matrix. In non-relativistic kinematics, the $S$-matrix is
\begin{equation}
S_{\alpha'\alpha} = 1 - 2\pi i
\sqrt{\mu_\alpha \mu_{\alpha'} k_\alpha k_{\alpha'}} \, T_{\alpha'\alpha}(E+i0^+;k_\alpha',k_\alpha) \,,
\end{equation}
where $k_\alpha$ and $\mu_\alpha$ are the on-shell momentum and reduced mass of channel $\alpha$.

The $T$-matrix is obtained by solving the coupled-channel Lippmann-Schwinger equation,
\begin{align}
T_{\beta'\beta}(z;p',p) &= V_{\beta'\beta}(p',p) +\sum_{\beta''}\int dq \, q^2 \, \nonumber \\
&
\hspace*{-0.50cm} \times V_{\beta'\beta''}(p',q) \frac{1}{z-E_{\beta''}(q)} T_{\beta''\beta}(z;q,p) \,,
\end{align}
where $\beta$ represents the set of quantum numbers necessary to determine a partial wave in the meson-meson channel, $V_{\beta'\beta}(p',p)$ is the full RGM potential, sum of direct and exchange kernels, and  $E_{\beta''}(q)$ is the energy for the momentum $q$ referred to the lower threshold.

Resonances correspond to poles in the complex energy plane,
\begin{equation}
\bar E = M_r - i\Gamma_r/2 \,,
\end{equation}
from which the mass $M_r$ and total width $\Gamma_r$ are extracted. Two Riemann sheets can be defined for each channel. The first Riemann sheet is defined for $0\le{\rm arg}(k_\alpha)<\pi$, and poles in this sheet are interpreted as bound states. The second Riemann sheet is defined for $\pi\le{\arg}(k_\alpha)<2\pi$) in such a way that poles in this region are identified as virtual or resonance states, depending if they lie below or above the threshold, respectively.

\subsection{Partial widths and branching ratios}

Some caution should be taken in order to obtain the partial widths of the resonances to a specific final meson-meson channel. Following Refs.~\cite{Ortega:2012rs, Grassi:2000dz}, in the neighborhood of a resonance, the $S$-matrix can be approximated as
\begin{align}
S_{\beta'\beta}(E) &= S_{\beta'\beta}^{bg}(E) - i 2\pi \delta^{(4)}(P_f-P_i) \dfrac{g_{\beta'}g_{\beta}}{E-\bar E} \,,
\end{align}
where $g_{\beta}$ are the residues of the pole, which can be interpreted as the amplitude of the resonance to the final state. Therefore, the partial width into a two-meson channel is
\begin{equation}
\hat\Gamma_\beta = 2\pi \, \frac{E_1E_2}{M_r} \, k_\beta |g_\beta|^2 \,.
\end{equation}
Since $\sum_\beta \hat\Gamma_\beta \neq \Gamma_r$ in general, branching ratios are defined as
\begin{equation}
{\cal B}_\beta = \frac{\hat\Gamma_\beta}{\sum_{\beta'} \hat\Gamma_{\beta'}} \,,
\end{equation}
and physical partial widths are obtained from
\begin{equation}
\Gamma_\beta = {\cal B}_\beta \, \Gamma_r \,,
\end{equation}
with $\Gamma_r=-2\,\text{Im}(\bar E)$.


\begin{table*}[t!]
\caption{\label{tab:channels} The meson-meson channels, with their mass thresholds in GeV/c$^2$ taken from Ref.~\cite{ParticleDataGroup:2024cfk}, considered in this work. We also show for each channel the included partial waves, denoted as $^{2S+1}L_J$, which are compatible with the different spin-parity ($J^P$) sectors.}
\begin{ruledtabular}
\begin{tabular}{cc|cccccc}
Channels & Thresholds & $0^-$ & $0^+$ & $1^-$ & $1^+$ & $2^-$ & $2^+$ \\
\hline
\tstrut
$\eta_b(1S)\eta_b(1S)$ & $18.797$ & $\cdots$ & ${}^1S_0$ & $\cdots$ & $\cdots$ & $\cdots$ & ${}^1D_2$ \\[0.5ex]
$\eta_b(1S)\Upsilon(1S)$ & $18.859$ & ${}^3P_0$ & $\cdots$ & ${}^3P_1$ & ${}^3S_1-{}^3D_1$ & ${}^3P_2$ & ${}^3D_2$ \\[0.5ex]
$\Upsilon(1S) \Upsilon(1S)$ & $18.921$ & ${}^3P_0$ & ${}^1S_0-{}^5D_0$ & ${}^3P_1$ & $\cdots$ & ${}^3P_2$ & ${}^5S_2-{}^1D_2-{}^5D_1$ \\[0.5ex]
$\eta_b(1S)\eta_b(2S)$ & $19.398$ & $\cdots$ & ${}^1S_0$ & ${}^1P_1$ &$ \cdots$ & $\cdots$ & ${}^1D_2$ \\[0.5ex]
$\eta_b(1S)\Upsilon(2S)$ & $19.422$ & ${}^3P_0$ & $\cdots$ & ${}^3P_1$ & ${}^3S_1-{}^3D_1$ & ${}^3P_2$ & ${}^3D_2$ \\[0.5ex]
$\eta_b(2S)\Upsilon(1S)$ & $19.459$ & ${}^3P_0$ & $\cdots$ & ${}^3P_1$ & ${}^3S_1-{}^3D_1$ & ${}^3P_2$ & ${}^3D_2$\\[0.5ex]
$\Upsilon(1S)\Upsilon(2S)$ & $19.484$& ${}^3P_0$ & ${}^1S_0-{}^5D_0$ & ${}^1P_1-{}^3P_1-{}^5P_1$ & ${}^3S_1-{}^3D_1$ & ${}^3P_2$ & ${}^5S_2-{}^1D_2-{}^3D_1-{}^5D_1$ \\[0.5ex]
$\eta_b(2S)\eta_b(2S)$ & $19.998$ & $\cdots$ & ${}^1S_0$ & $\cdots$  & $\cdots$ & $\cdots$ & ${}^1D_2$ \\[0.5ex]
$\eta_b(2S)\Upsilon(2S)$ & $20.022$ & ${}^3P_0$ & $\cdots$ & ${}^3P_1$ & ${}^3S_1-{}^3D_1$ & ${}^3P_2$ & ${}^3D_2$ \\[0.5ex]
$\Upsilon(2S)\Upsilon(2S)$ & $20.047$ & ${}^3P_0$ & ${}^1S_0-{}^5D_0$ & ${}^3P_1$ & $\cdots$ & ${}^3P_2$ & ${}^5S_2-{}^1D_2-{}^5D_1$ \\
\end{tabular}
\end{ruledtabular}
\end{table*}


\begin{turnpage}
\begin{table*}[!t]
\caption{\label{tab:Tuu1} Coupled-channels calculation of the $J^P=0^\pm$, $1^\pm$ and $2^\pm$ $bb \bar b \bar b$ sectors ($T_{\Upsilon\Upsilon}$ states) as meson-meson molecules, including the channels detailed in Table~\ref{tab:channels}. Errors are estimated by varying the strength of the potential by $\pm10\%$. To simplify the notation, we denote $\eta_b$, $\eta_b^\prime$, $\Upsilon$ and $\Upsilon^\prime$ for $\eta_b(1S)$, $\eta_b(2S)$, $\Upsilon(1S)$ and $\Upsilon(2S)$, respectively. \emph{$1^{st}$ column:} Pole's quantum numbers; \emph{$2^{nd}$ column:} Pole's mass in GeV/c$^2$; \emph{$3^{rd}$ column:} Pole's width in MeV; \emph{$4^{th}$-$13^{th}$ columns:} $(b\bar b)-(b\bar b)$ wavefunction composition attending to the meson-meson channel probability, in \%. States with a dagger before their mass are virtual states, defined as poles in the second Riemann sheet below their closest threshold.}
\begin{ruledtabular}
\begin{tabular}{lllllllllllll}
$J^{PC}$ & $M_{\text{pole}}$ & $\Gamma_{\text{pole}}$  & ${\cal P}_{\eta_b\eta_b}$ & ${\cal P}_{\eta_b\Upsilon}$ & ${\cal P}_{\Upsilon\Upsilon}$ & ${\cal P}_{\eta_b\eta_b^\prime}$ & ${\cal P}_{\eta_b\Upsilon^\prime}$ & ${\cal P}_{\eta_b^\prime\Upsilon}$ & ${\cal P}_{\Upsilon\Upsilon^\prime}$ & ${\cal P}_{\eta_b^\prime\eta_b^\prime}$ & ${\cal P}_{\eta_b^\prime\Upsilon^\prime}$ & ${\cal P}_{\Upsilon^\prime\Upsilon^\prime}$ \\
\hline
\tstrut
$ 0^{--}$ & $19.546_{-3}^{+2}$ & $312_{-15}^{+16}$  & 0  & 0  & 0  & 0  & $54.85 \pm 0.04$  & $45.04_{-0.03}^{+0.04}$  & 0  & 0  & $0.058 \pm 0.005$  & 0 \\[2ex]
$ 1^{--}$ & $19.546_{-3}^{+2}$ & $311_{-15}^{+16}$  & 0  & 0  & 0  & 0  & $54.84 \pm 0.04$  & $45.04 \pm 0.04$  & 0  & 0  & $0.058_{-0.005}^{+0.004}$  & 0 \\[2ex]
$ 2^{--}$ & $19.546_{-3}^{+2}$ & $311_{-15}^{+16}$  & 0  & 0  & 0  & 0  & $54.84 \pm 0.04$  & $45.04 \pm 0.04$  & 0  & 0  & $0.058_{-0.005}^{+0.004}$  & 0 \\[2ex]
$ 0^{++}$ & $18.894_{-3}^{+2}$ & $165_{-15}^{+17}$  & $70_{-2}^{+1}$  & 0  & $28 \pm 2$  & $1.2_{-0.2}^{+0.3}$  & 0  & 0  & $0.39_{-0.03}^{+0.05}$  & $0.5_{-0.1}^{+0.2}$  & 0  & $0.29_{-0.04}^{+0.06}$ \\
& $18.996 \pm 1$ & $119_{-6}^{+7}$  & $28 \pm 1$  & 0  & $71 \pm 1$  & $0.049_{-0.001}^{+0.005}$  & 0  & 0  & $0.49_{-0.06}^{+0.08}$  & $0.008 \pm 0.001$  & 0  & $0.13_{-0.02}^{+0.04}$  \\
& $19.443_{-3}^{+4}$ & $72 \pm 14$  & $4.2 \pm 0.8$  & 0  & $4 \pm 1$  & $74.1_{-0.4}^{+0.3}$  & 0  & 0  & $16.6_{-0.9}^{+0.2}$  & $0.5 \pm 0.1$  & 0  & $0.254_{-0.006}^{+0.008}$  \\
& $19.501 \pm 1$ & $32_{-9}^{+10}$  & $0.84_{-0.07}^{+0.05}$  & 0  & $10_{-1}^{+2}$  & $41 \pm 6$  & 0  & 0  & $49_{-8}^{+7}$  & $0.048_{-0.007}^{+0.008}$  & 0  & $0.24 \pm 0.04$  \\
& $19.900_{-11}^{+21}$ & $419_{-2}^{+13}$  & $3.0_{-0.2}^{+0.0}$  & 0  & $15.4_{-0.3}^{+0.1}$  & $6.0_{-0.4}^{+0.3}$  & 0  & 0  & $21 \pm 2$  & $12.0_{-0.6}^{+0.2}$  & 0  & $42 \pm 2$  \\
& $20.281_{-7}^{+10}$ & $283_{-5}^{+14}$  & $1.4 \pm 0.4$  & 0  & $7 \pm 2$  & $3.9_{-0.5}^{+0.1}$  & 0  & 0  & $17_{-3}^{+2}$  & $9_{-1}^{+2}$  & 0  & $61_{-3}^{+4}$ \\[2ex]
$ 1^{+-}$ & $18.948_{-2}^{+1}$ & $148_{-12}^{+14}$  & 0  & $98.2_{-0.4}^{+0.3}$  & 0  & 0  & $0.5 \pm 0.1$  & $0.7 \pm 0.1$  & 0  & 0  & $0.5_{-0.1}^{+0.2}$  & 0 \\
& $19.452 \pm 3$ & $130 \pm 12$  & 0  & $9.4 \pm 0.9$  & 0  & 0  & $51 \pm 1$  & $38.4_{-0.1}^{+0.0}$  & 0  & 0  & $1.2_{-0.2}^{+0.3}$  & 0  \\
& $19.474 \pm 1$ & $71_{-9}^{+10}$  & 0  & $7.1 \pm 0.7$  & 0  & 0  & $56 \pm 2$  & $36.8_{-0.9}^{+0.7}$  & 0  & 0  & $0.47_{-0.05}^{+0.07}$  & 0 \\
& $19.882_{-6}^{+7}$ & $551_{-25}^{+27}$  & 0  & $15.6_{-0.4}^{+0.5}$  & 0  & 0  & $16.55_{-0.02}^{+0.05}$  & $16.3_{-0.1}^{+0.0}$  & 0  & 0  & $51.5_{-0.4}^{+0.6}$  & 0  \\
& $20.261 \pm 1$ & $310_{-7}^{+9}$  & 0  & $5.1_{-0.0}^{+0.1}$  & 0  & 0  & $4.4_{-0.3}^{+0.4}$  & $5.1_{-0.4}^{+0.6}$  & 0  & 0  & $85.5_{-1.0}^{+0.6}$  & 0  \\
& $20.332_{-7}^{+6}$ & $480_{-9}^{+10}$  & 0  & $0.22_{-0.01}^{+0.02}$  & 0  & 0  & $15 \pm 1$  & $10 \pm 1$  & 0  & 0  & $75 \pm 2$  & 0  \\[2ex]
$ 2^{++}$ & $19.005_{-3}^{+2}$ & $176_{-12}^{+14}$  & 0  & 0  & $96.7_{-0.7}^{+0.5}$  & 0  & 0  & 0  & $2.3_{-0.3}^{+0.4}$  & 0  & 0  & $1.0_{-0.2}^{+0.3}$  \\
& $19.515 \pm 1$ & $84_{-9}^{+10}$  & 0  & 0  & $9.8 \pm 0.6$  & 0  & 0  & 0  & $89.5 \pm 0.7$  & 0  & 0  & $0.7 \pm 0.1$  \\
& $19.933_{-5}^{+4}$ & $576_{-20}^{+21}$  & 0  & 0  & $14.9 \pm 0.3$  & 0  & 0  & 0  & $30.6 \pm 0.1$  & 0  & 0  & $54.5 \pm 0.4$  \\
& $20.278 \pm 1$ & $321_{-8}^{+9}$  & 0  & 0  & $6.6_{-0.2}^{+0.1}$  & 0  & 0  & 0  & $8.4 \pm 0.1$  & 0  & 0  & $85.1_{-0.2}^{+0.0}$  \\
& \hspace*{-0.14cm}$^\dagger$$20.040 \pm 1$ & $38 \pm 6$  & 0  & 0  & $1.0 \pm 0.2$  & 0  & 0  & 0  & $6.8_{-0.7}^{+0.6}$  & 0  & 0  & $92.2 \pm 0.9$ \\
\end{tabular}
\end{ruledtabular}
\end{table*}
\end{turnpage}

\begin{table*}[!t]
\caption{\label{tab:Tuu2} Coupled-channels calculation of the $J^P=0^\pm$, $1^\pm$ and $2^\pm$ $bb \bar b \bar b$ sectors ($T_{\Upsilon\Upsilon}$ states) as meson-meson molecules, including the channels detailed in Table~\ref{tab:channels}. Errors are estimated by varying the strength of the potential by $\pm10\%$. To simplify the notation, we denote $\eta_b$, $\eta_b^\prime$, $\Upsilon$ and $\Upsilon^\prime$ for $\eta_b(1S)$, $\eta_b(2S)$, $\Upsilon(1S)$ and $\Upsilon(2S)$, respectively.  \emph{$1^{st}$ column:} Pole's quantum numbers; \emph{$2^{nd}$ column:} Pole's mass in GeV/c$^2$; \emph{$3^{rd}$ column:} Pole's width in MeV; \emph{$4^{th}$-$13^{th}$ columns:} Branching ratios in \%. States with a dagger before their mass are virtual states, defined as poles in the second Riemann sheet below their closest threshold.}
\begin{ruledtabular}
\begin{tabular}{lllllllllllll}
$J^{PC}$ & $M_{\text{pole}}$ & $\Gamma_{\text{pole}}$  & ${\cal B}_{\eta_b\eta_b}$ & ${\cal B}_{\eta_b\Upsilon}$ & ${\cal B}_{\Upsilon\Upsilon}$ & ${\cal B}_{\eta_b\eta_b^\prime}$ & ${\cal B}_{\eta_b\Upsilon^\prime}$ & ${\cal B}_{\eta_b^\prime\Upsilon}$ & ${\cal B}_{\Upsilon\Upsilon^\prime}$ & ${\cal B}_{\eta_b^\prime\eta_b^\prime}$ & ${\cal B}_{\eta_b^\prime\Upsilon^\prime}$ & ${\cal B}_{\Upsilon^\prime\Upsilon^\prime}$ \\
\hline
\tstrut
$ 0^{--}$ & $19.546_{-3}^{+2}$ & $312_{-15}^{+16}$  & 0  & 0  & 0  & 0  & $54.0_{-0.3}^{+0.2}$  & $46.0 \pm 0.3$  & 0  & 0  & 0  & 0  \\[2ex]
$ 1^{--}$ & $19.546_{-3}^{+2}$ & $311_{-15}^{+16}$  & 0  & 0  & 0  & 0  & $54.0 \pm 0.3$  & $46.0 \pm 0.3$  & 0  & 0  & 0  & 0  \\[2ex]
$ 2^{--}$ & $19.546_{-3}^{+2}$ & $311_{-15}^{+16}$  & 0  & 0  & 0  & 0  & $53.9_{-0.2}^{+0.4}$  & $46.0_{-0.3}^{+0.2}$  & 0  & 0  & 0  & 0  \\[2ex]
$ 0^{++}$ & $18.894_{-3}^{+2}$ & $165_{-15}^{+17}$  & $100 \pm 0$  & 0  & 0  & 0  & 0  & 0  & 0  & 0  & 0  & 0  \\
& $18.996 \pm 1$ & $119_{-6}^{+7}$  & $39_{-3}^{+2}$  & 0  & $61_{-2}^{+3}$  & 0  & 0  & 0  & 0  & 0  & 0  & 0  \\
& $19.443_{-3}^{+4}$ & $72 \pm 14$  & $26_{-2}^{+1}$  & 0  & $14_{-3}^{+4}$  & $60 \pm 2$  & 0  & 0  & 0  & 0  & 0  & 0  \\
& $19.501 \pm 1$ & $32_{-9}^{+10}$  & $6.4_{-0.2}^{+0.3}$  & 0  & $40 \pm 2$  & $24 \pm 2$  & 0  & 0  & $29_{-4}^{+3}$  & 0  & 0  & 0  \\
& $19.900_{-11}^{+21}$ & $419_{-2}^{+13}$  & $12_{-1}^{+0}$  & 0  & $65_{-3}^{+4}$  & $10 \pm 1$  & 0  & 0  & $13_{-2}^{+3}$  & 0  & 0  & 0  \\
& $20.281_{-7}^{+10}$ & $283_{-5}^{+14}$  & $1_{-0}^{+1}$  & 0  & $10_{-4}^{+7}$  & $6_{-2}^{+1}$  & 0  & 0  & $17 \pm 7$  & $3 \pm 3$  & 0  & $62_{-2}^{+1}$  \\[2ex]
$ 1^{+-}$ & $18.948_{-2}^{+1}$ & $148_{-12}^{+14}$  & 0  & $100 \pm 0$  & 0  & 0  & 0  & 0  & 0  & 0  & 0  & 0  \\
& $19.452 \pm 3$ & $130 \pm 12$  & 0  & $42.9_{-0.7}^{+0.5}$  & 0  & 0  & $57.1_{-0.5}^{+0.7}$  & 0  & 0  & 0  & 0  & 0  \\
& $19.474.4 \pm 1$ & $71_{-9}^{+10}$  & 0  & $40 \pm 1$  & 0  & 0  & $37.6_{-0.1}^{+0.3}$  & $22 \pm 1$  & 0  & 0  & 0  & 0  \\
& $19.882_{-6}^{+7}$ & $551_{-25}^{+27}$  & 0  & $62_{-4}^{+3}$  & 0  & 0  & $19 \pm 1$  & $18_{-2}^{+3}$  & 0  & 0  & 0  & 0  \\
& $20.261 \pm 1$ & $310_{-7}^{+9}$  & 0  & $31_{-1}^{+0}$  & 0  & 0  & $0.2_{-0.1}^{+0.2}$  & $1.0_{-0.5}^{+0.9}$  & 0  & 0  & $68_{-1}^{+2}$  & 0  \\
& $20.332_{-7}^{+6}$ & $480_{-9}^{+10}$  & 0  & 0  & 0  & 0  & $1.7 \pm 0.2$  & $1.7 \pm 0.2$  & 0  & 0  & $96.6 \pm 0.4$  & 0  \\[2ex]
$ 2^{++}$ & $19.005_{-3}^{+2}$ & $176_{-12}^{+14}$  & 0  & 0  & $100 \pm 0$  & 0  & 0  & 0  & 0  & 0  & 0  & 0  \\
& $19.515 \pm 1$ & $84_{-9}^{+10}$  & 0  & 0  & $37 \pm 1$  & 0  & 0  & 0  & $63 \pm 1$  & 0  & 0  & 0  \\
& $19.933_{-5}^{+4}$ & $576_{-20}^{+21}$  & 0  & 0  & $50 \pm 4$  & 0  & 0  & 0  & $50 \pm 4$  & 0  & 0  & 0  \\
& $20.278 \pm 1$ & $321_{-8}^{+9}$  & 0  & 0  & $29_{-3}^{+2}$  & 0  & 0  & 0  & 0  & 0  & 0  & $71 \pm 2$  \\
& \hspace*{-0.14cm}$^\dagger$$20.040 \pm 1$ & $38 \pm 6$  & 0  & 0  & $30 \pm 5$  & 0  & 0  & 0  & $70 \pm 5$  & 0  & 0  & 0  \\
\end{tabular}
\end{ruledtabular}
\end{table*}

\section{RESULTS}
\label{sec:results}

In this section we present the results of the coupled-channels calculation of the $b\bar b - b\bar b$ system in the $J^P = 0^\pm, 1^\pm, 2^\pm$ sectors.
The $C$-parity of a meson–meson pair is defined as the product of the intrinsic $C$-parities of the individual bottomonium states. Explicitly, $C=(-1)^{L_A+S_A}\cdot (-1)^{L_B+S_B}$, which evaluates to $C = +1$ for $PP$ and $VV$ channels, and $C = -1$ for $PV$ channels, where $P$ denotes a pseudoscalar meson and $V$ a vector meson. The channels and partial waves included are listed in Table~\ref{tab:channels} and correspond to all combinations of the lowest $S$-wave bottomonia, namely $\Upsilon(1S)$, $\Upsilon(2S)$, $\eta_b(1S)$, and $\eta_b(2S)$. We restrict ourselves to relative orbital angular momenta $L \le 2$, since higher partial waves are expected to play a negligible role in the energy region considered.

As discussed in Sec.~\ref{sec:theory}, direct interactions between two color-singlet bottomonia are suppressed. The confinement potential does not generate a direct contribution, and gluon-annihilation diagrams are small in the heavy-quark sector. Consequently, the dominant interaction arises from quark-exchange mechanisms. This feature implies that the resulting states cannot be interpreted as loosely bound systems driven by residual meson-meson forces alone; instead, they correspond to compact meson-meson configurations generated by short-range quark rearrangement. Nevertheless, throughout this work we refer to them generically as molecular states, in the sense of resonant structures emerging from the interaction of two color-singlet mesons.

Before discussing the spectrum, it is worth recalling that the model parameters are constrained by a global fit to hadron observables within an accuracy of about $10-20\%$. This induces a theoretical uncertainty in the pole positions, and related physical observables. To estimate its impact, we vary the strength of the potentials by $\pm 10\%$ and propagate the changes as theoretical uncertainties in the pole properties.

The results of the coupled-channels calculation are summarized in Table~\ref{tab:Tuu1} (pole positions and wave-function compositions) and Table~\ref{tab:Tuu2} (masses, widths, and branching ratios). We find a total of $20$ poles distributed across the different $J^P$ sectors: $1$ in $0^{--}$, $6$ in $0^{++}$, $1$ in $1^{--}$, $6$ in $1^{+-}$, $1$ in $2^{--}$, and $5$ in $2^{++}$. Their masses span the region between the lowest threshold, $\eta_b(1S)\eta_b(1S)$, and the highest included one, $\Upsilon(2S)\Upsilon(2S)$, while the widths range from a few tens of MeV up to several hundred MeV.

A clear manifestation of heavy-quark spin symmetry (HQSS) is observed in the approximate degeneracy among states belonging to the $\{0^{--},1^{--},2^{--}\}$ and $\{0^{++},1^{+-},2^{++}\}$ multiplets. In particular, the three poles found around $M \simeq 19.55$ GeV in the $0^{--}$, $1^{--}$, and $2^{--}$ sectors exhibit nearly identical masses, widths, and channel compositions, being dominated by the $\eta_b(1S)\Upsilon(2S)$ and $\eta_b(2S)\Upsilon(1S)$ components, which form the $^3P_J$ triplet with $J=\{0,1,2\}$. Similar HQSS patterns appear in the positive-parity sectors due to the ${}^1S_0-{}^3S_1-{}^5S_2$ symmetry, although sizable splittings arise due to the different partial waves that contribute in each case.

The wave-function decomposition in Table~\ref{tab:Tuu1} reveals that many states are strongly dominated by a single meson-meson channel near threshold, such as the $0^{++}$ state at $M \simeq 18.89$ GeV, which is dominated by $\eta_b(1S)\eta_b(1S)$, the $1^{+-}$ state at $\sim\!\!18.95$ GeV, which is almost entirely $\eta_b(1S)\Upsilon(1S)$, or the $2^{++}$ state around $19.01$ GeV, largely dominated by $\Upsilon(1S)\Upsilon(1S)$. These structures can be interpreted as threshold-driven resonances. In contrast, higher-mass states generally display a more fragmented structure, with significant admixtures of channels involving radially excited bottomonia.

Regarding decay properties, the results shown in Table~\ref{tab:Tuu2} indicate that searches limited to the $\eta_b(1S)\eta_b(1S)$, $\eta_b(1S)\Upsilon(1S)$, and $\Upsilon(1S)\Upsilon(1S)$ channels would probe only a restricted portion of the predicted spectrum, spanning the mass range from $18.89$ to $19.90$ GeV. The branching ratios reported in Table~\ref{tab:Tuu2} further reveal that many resonances couple strongly to channels containing at least one excited bottomonium state. This behavior is particularly evident in the negative-parity sectors, where the only non-zero branching fractions correspond to the $\eta_b(1S)\Upsilon(2S)$ and $\eta_b(2S)\Upsilon(1S)$ final states. A similar pattern is observed for several $0^{++}$, $1^{+-}$, and $2^{++}$ states above $20$ GeV, which predominantly decay into $\Upsilon(1S)\Upsilon(2S)$, $\eta_b(2S)\Upsilon(2S)$, or $\Upsilon(2S)\Upsilon(2S)$. These results suggest that experimental searches focusing exclusively on pairs of ground-state bottomonia may overlook a significant fraction of the spectrum. In contrast, channels such as $\eta_b(1S)\Upsilon(2S)$, $\eta_b(2S)\Upsilon(1S)$, $\Upsilon(1S)\Upsilon(2S)$, and $\Upsilon(2S)\Upsilon(2S)$ emerge as particularly promising discovery modes. Therefore, final states involving mixed bottomonium excitations should be considered a primary target in future experimental analyses.

Overall, the spectrum of $T_{\Upsilon\Upsilon}$ candidates displays the same qualitative features found in the fully charmed sector: a rich pattern of resonant structures generated by coupled-channel dynamics, approximate HQSS multiplets, and strong correlations between pole positions and nearby thresholds. However, the larger bottom mass shifts the whole spectrum upward and modifies phase space and partial-wave effects, leading in general to broader states and stronger mixing among channels with excited bottomonia.

From the experimental point of view, the observation of the $T_{\Upsilon\Upsilon}$ spectrum is clearly more challenging than in the charmonium sector due to the higher masses and the correspondingly smaller production cross sections. Nevertheless, the clean leptonic decays of the $\Upsilon(nS)$ states provide very favorable experimental signatures. In this context, high-luminosity hadron facilities such as LHCb, CMS, and ATLAS offer the most promising environment for double-bottomonium spectroscopy.

Although the detection of $\eta_b$ mesons is experimentally more demanding, several of the predicted states couple strongly to $\eta_b(1S)\Upsilon(2S)$ and $\eta_b(2S)\Upsilon(1S)$. Indirect access to these channels may be possible through radiative or hadronic transitions of the excited bottomonia, which could provide complementary information on the structure of the resonances. In particular, correlations between structures observed in $\Upsilon(1S)\Upsilon(2S)$ and in channels involving $\eta_b$ could help to disentangle HQSS partner states.

Finally, the presence of broad resonances with sizable couplings to multiple channels suggests that the experimental signal may appear as enhancements or distortions in invariant-mass distributions rather than as narrow peaks. A coupled-channel amplitude analysis, incorporating simultaneously ground and excited bottomonia, would be the most suitable strategy to reveal the underlying pole structure. Such measurements would provide a stringent test of the molecular picture driven by quark-exchange dynamics and of heavy-quark spin symmetry in fully heavy tetraquark systems.


\section{SUMMARY}
\label{sec:summary}

We have performed a coupled-channels study of the $b\bar b - b\bar b$ system including all combinations of the lowest $S$-wave bottomonia: $\eta_b(1S)$, $\eta_b(2S)$, $\Upsilon(1S)$ and $\Upsilon(2S)$; in the $J^P = 0^\pm, 1^\pm, 2^\pm$ sectors. The meson-meson interaction is derived from a constituent quark model via the resonating group method, and the properties of the $T_{\Upsilon\Upsilon}$ candidates are extracted as poles of the scattering matrix.

The calculation predicts a rich spectrum of states between the lowest and highest considered thresholds. A total of $20$ poles are found in the different spin-parity sectors, including both resonant and virtual states. Their widths cover a wide range, from relatively narrow structures to broad resonances strongly coupled to multiple channels.

A clear imprint of heavy-quark spin symmetry is observed in the approximate degeneracy of several $\{0^{--},1^{--},2^{--}\}$ and $\{0^{++},1^{+-},2^{++}\}$ multiplets, whose members share similar masses, widths, and channel compositions. Many of the predicted states are closely correlated with nearby meson-meson thresholds and are dominated by a single channel, indicating a strong threshold-driven dynamics. At higher energies, mixing among channels with radially excited bottomonia becomes increasingly important.

The analysis of branching ratios shows that a substantial fraction of the spectrum decays predominantly into channels involving at least one excited bottomonium, particularly $\Upsilon(1S)\Upsilon(2S)$, $\Upsilon(2S)\Upsilon(2S)$, and channels with $\eta_b(2S)$. Consequently, experimental searches restricted to meson-meson final states involving $\eta_b(1S)$ and $\Upsilon(1S)$ may access only a limited part of the spectrum.

Overall, the $T_{\Upsilon\Upsilon}$ spectrum exhibits the same qualitative features previously found in the fully charmed sector, namely the emergence of multiple resonant structures from coupled-channel dynamics, approximate heavy-quark spin symmetry multiplets, and strong correlations between pole positions and thresholds. The present results offer concrete predictions for masses, widths, and dominant decay modes, and provide a framework to interpret future measurements of double-bottomonium invariant-mass spectra at high-luminosity hadron facilities.


\begin{acknowledgments}
This work has been partially funded by
EU Horizon 2020 research and innovation program, STRONG-2020 project, under grant agreement no. 824093 and Ministerio Espa\~nol de Ciencia e Innovaci\'on, grant nos. PID2019-105439GB-C22 and PID2022-140440NB-C22.
\end{acknowledgments}


\bibliography{print_Tupsups}

\end{document}